\documentclass[reprint,preprintnumbers,
amsmath,amssymb,aps,nofootinbib,superscriptaddress]{revtex4-1}
\usepackage[pdftex]{graphicx}
\usepackage{dcolumn}
\usepackage{natbib}
\usepackage[pdftex]{hyperref}
\usepackage{color}
\definecolor{phthaloblue}{rgb}{0.0, 0.06, 0.54}
\definecolor{bluscuro}{rgb}{0.15, 0.2, .85}
\definecolor{rossos}{cmyk}{0,1,1,0.55}
\definecolor{bluchiaro}{cmyk}{1,.3,0.,0.1}
\definecolor{ultramarine}{rgb}{0.07, 0.04, 0.56}
\definecolor{cadmiumgreen}{rgb}{0.0, 0.42, 0.24}
\definecolor{indigo(dye)}{rgb}{0.0, 0.25, 0.42}
\hypersetup{
    colorlinks=true,
    linkcolor=ultramarine,
    citecolor=ultramarine,
    filecolor=indigo(dye),
    urlcolor=indigo(dye),
    }

\newcommand{\para}[1]{\par\vspace{2mm}\noindent\textbf{\emph{{#1}}.---}}

\begin{document}
\preprint{\begin{minipage}[b]{1\linewidth}
\begin{flushright} TU-1068, IPMU18-0132, MIT-CTP/5040\end{flushright}
\end{minipage}}

\title{Eternal Inflation and 
Swampland Conjectures}
\author{Hiroki Matsui}
\email{hiroki.matsui.c6@tohoku.ac.jp}
\affiliation{Department of Physics, Tohoku University, Sendai, Miyagi 980-8578, Japan}
\author{Fuminobu Takahashi}
\email{fumi@tohoku.ac.jp}
\affiliation{Department of Physics, Tohoku University, Sendai, Miyagi 980-8578, Japan}\affiliation{Kavli Institute for the Physics and Mathematics of the Universe (Kavli IPMU), UTIAS, WPI, The University of Tokyo, Kashiwa, Chiba 277-8568, Japan}	
\affiliation{Center for Theoretical Physics, Massachusetts Institute of Technology, Cambridge, MA 02139, U.S.A.}

%%%%%%%%%%%%%%%%%%%%%%%%%%%%%%%%%%%%%%%%%%%%%%%%%%%%%%%%%%%%%%%  

\begin{abstract}
We study if eternal inflation is realized while satisfying the recently proposed string Swampland
criteria concerning the range of scalar field excursion, $|\Delta \phi| < \mathcal{D} \cdot M_{\rm P}$, and the potential gradient, 
$|\nabla V| > c \cdot V/M_{\rm P}$, where $\mathcal{D}$ and $c$ are constants of order unity, and $M_{\rm P}$ is the reduced Planck mass.
We find that only the eternal inflation of chaotic type is possible for $c \sim {\cal O}(0.01)$ and $1/\mathcal{D} \sim  {\cal O}(0.01)$, and
that the Hubble parameter during the eternal inflation is
parametrically close to the Planck scale, and is
 in the range of $2 \pi c \lesssim H_{\rm inf}/M_{\rm P} < 1/\sqrt{3}$.
\end{abstract}
\maketitle

%%%%%%%%%%%%%%%%%%%%%%%%%%
\section{Introduction}
\label{sec:Swampland}
%%%%%%%%%%%%%%%%%%%%%%%%%%
The inflation is a quasi de Sitter expansion of the Universe~\cite{Starobinsky:1980te,Guth:1980zm,Sato:1980yn,Linde:1981mu,Albrecht:1982wi}, which not only solves various initial condition problems of the standard
big bang cosmology, but also provides a natural explanation of the origin of density perturbations. 
Accumulating observational evidences support that our Universe experienced inflation at an
early stage of the evolution (see e.g. Ref.~\cite{Akrami:2018odb}).

When the currently observable scales exited the horizon during inflation, 
the energy scale was much smaller than the Planck scale. Therefore, the inflation can be well described 
by a low-energy effective theory. This does not necessarily mean, however, that any inflation models realized in the language of the effective theory can be UV completed.
In fact, it is widely believed that consistent-looking effective field theories could be 
in the swampland that cannot be consistently embedded in quantum gravity~\cite{Vafa:2005ui}.
It is of utmost importance to have the correct criteria to identify the boundary between
the landscape and the swampland.

There have been proposed several conjectures on such criteria motivated by
 black hole physics~\cite{ArkaniHamed:2006dz} or 
 string compactifications~\cite{Ooguri:2016pdq,Freivogel:2016qwc,Brennan:2017rbf}. 
Recently it was proposed that an effective field theory that can be embedded consistently in
quantum gravity must satisfy the two conditions concerning the 
range of scalar field excursion and the potential gradient~\cite{Ooguri:2006in,Obied:2018sgi}.\footnote{
We emphasize here that the conditions are still conjectures that have not been proven to be correct yet.
In fact, there are many papers that criticize the conjectures~\cite{Cicoli:2018kdo,Kachru:2018aqn,Kallosh:2018psh,Kallosh:2018wme,Akrami:2018ylq}.
}
These conjectures are known to tightly constrain possible inflation theories, especially those based
on a single-field slow-roll inflation regime. This lends support to
more complicated inflation models or non-standard history of the Universe such as
chromo-nautral inflation~\cite{Agrawal:2018mkd}, multi-field inflation~\cite{Achucarro:2018vey}
or curvaton scenarios~\cite{Kehagias:2018uem}.

In this paper we study if eternal inflation is realized while satisfying the above two swampland criteria.
The eternal inflation~\cite{1983veu..conf..251S,Vilenkin:1983xq,Linde:1986fc,Linde:1986fd,
Goncharov:1987ir,Guth:2007ng} precedes the inflation epoch during which the currently observable
scales exited the horizon, and it is considered to play an important role in populating various vacua 
(and therefore various theoretical possibilities) 
in the landscape. The eternal inflation can even explain a contrived fine-tuning of the parameters or setup that is necessary for the complicated life to emerge based on the anthropic argument.
We will show in the later sections that only the eternal inflation of chaotic type is 
possible, and it requires that the numerical coefficients 
appearing in the swampland criteria are parametrically smaller or larger than unity.

%%%%%%%%%%%%%%%%%%%%%%%%%%
\section{Swampland Distance/de Sitter Conjectures }
\label{sec:swamplandc}
%%%%%%%%%%%%%%%%%%%%%%%%%%
Here we briefly summarize the recently proposed two swampland criteria. 
The first one is the Swampland Distance Conjecture~\cite{Ooguri:2006in,Klaewer:2016kiy},
which states that any effective field theories are valid only for 
a finite variation of the scalar field, $\Delta \phi$.
In the string compactification, the mass of an infinite tower of 
states (collectively denoted by $m$) become exponentially light in the field distance $\Delta\phi$, 
\begin{align}
m^2\sim \exp \left( -\frac{\Delta\phi}{\mathcal{D}M _{\rm P }} \right) , 
\end{align}
where  ${M}_{\rm P } \simeq 2.4 \times 10^{18}$\,GeV is the reduced Planck mass, and 
$\mathcal{D}$ is a numerical constant of order unity, but its
precise value depends on details of the compactification. 
 One can see that
the large field deviation, $\Delta\phi \gtrsim \mathcal{D} M _{\rm P}$, would spoil the low-energy effective field theories,
since a tower of the heavy states continuously become so light that they generally affect the low-energy
dynamics. 
Thus, the following condition must be satisfied for the validity of the low-energy description,
\begin{align}
\frac{\left|\Delta m^2\right|}{m^2}\lesssim {\cal O}(1)\quad\Rightarrow\quad \Delta\phi 
\lesssim \mathcal{D}M _{\rm P } \quad,
\end{align}
where the first inequality is such that heavy states should remain heavy during the scalar field excursion, $\Delta\phi$.
For $\mathcal{D} = {\cal O}(1)$, the field variation $\Delta\phi$ cannot be much larger than 
the Planck mass.
See e.g. Refs.~\cite{
Klaewer:2016kiy,Blumenhagen:2017cxt,Palti:2017elp,Grimm:2018ohb,
Heidenreich:2018kpg} for evidences supporting the conjecture.

The second one is the Swampland de Sitter Conjecture~\cite{Obied:2018sgi}, which
states that the low-energy effective potential, $V$,
consistent with quantum gravity must satisfy
\begin{align}
| \nabla V | > c\, \frac{V}{M _{\rm P }} \label{eq:conj2} \,,
\end{align}
where $| \nabla V |$ is the norm of the gradient of $V$
on the scalar manifold, and $c$ is a  constant. 
The conjecture is based on recent discussions and criticisms
about constructing de Sitter vacua in string theory
(see e.g.~\cite{Danielsson:2018ztv} for a recent review).
The bound (\ref{eq:conj2}) forbids de Sitter vacuum solutions 
and also restricts possible slow-roll inflation scenarios. 
The validity of these 
conjectures and their cosmological implications have been recently 
discussed in Refs.~\cite{Hertzberg:2007wc,Agrawal:2018own,Andriot:2018wzk,
Dvali:2018fqu,Landete:2018kqf,Banerjee:2018qey,Aalsma:2018pll,Achucarro:2018vey,
Garg:2018reu,Kehagias:2018uem,Lehners:2018vgi,
Dias:2018ngv,Denef:2018etk,Colgain:2018wgk,2018arXiv180709538R,
Roupec:2018mbn,Andriot:2018ept,Cicoli:2018kdo,Kachru:2018aqn,Kallosh:2018psh,
Kallosh:2018wme,Akrami:2018ylq}.

The condition for inflation is formally written as
\begin{align}
\frac { \ddot { a }  }{ a } ={ H }^{ 2 }\left( 1-\epsilon  \right)>0\quad\Rightarrow\quad 
 \epsilon\equiv \frac { -\dot { H }  }{ H^2 }<1\,,
\end{align}
where $a(t)$ is the scale factor, the overdot denotes the derivative with respect to the cosmic time, $t$,
and $H(t) = \dot{a}/a$ is the Hubble parameter.
The slow-roll condition requires $\epsilon <1$, while  
the conjecture (\ref{eq:conj2}) restricts the slow-roll parameter,
\begin{align}
\epsilon \simeq \frac{M_{\rm P}^2}{2} \left(\frac{\nabla V}{V}\right)^2
\quad\Rightarrow\quad 
\epsilon^{ 1/2 } >\frac { c }{ \sqrt { 2 }}.
\label{eq:epsilonc}
\end{align}
Thus, the slow-roll inflation requires $c<  \sqrt { 2 }$.

In fact, the Planck measurements 
of the Cosmic Microwave Background (CMB)~\cite{Akrami:2018odb} further restricts 
the slow-roll parameter as $\epsilon <0.0045$,
which leads to $c < 0.094$. Therefore, the
condition (\ref{eq:conj2}) cannot be satisfied
if $c$ is strictly equal to unity in a simple inflation model. 
Note that more complicated inflation models or non-standard history of the Universe such as 
multi-field inflation~\cite{Achucarro:2018vey} or curvaton-like scenarios~\cite{Kehagias:2018uem}
can ameliorate the tension.

Lastly we emphasize that the precise values of $c$ and $\mathcal{D}$ depend on the detailed 
string construction and they can be slightly larger or smaller than unity~\cite{Dias:2018ngv}. 
In the following sections, we study if eternal inflation can be realized while satisfying the above two
swampland conjectures. In particular, we derive conditions on $c$ and $\mathcal{D}$, allowing
possible deviations from their canonical values.

%%%%%%%%%%%%%%%%%%%%%%%%%%
\section{Eternal Inflation }
\label{sec:EternalInflation}
%%%%%%%%%%%%%%%%%%%%%%%%%%
One of the most intriguing properties of the inflationary paradigm is that
inflation can be eternal~\cite{1983veu..conf..251S,Vilenkin:1983xq,Linde:1986fc,Linde:1986fd,
Goncharov:1987ir} (see also Ref.~\cite{Guth:2007ng}).
 Once the eternal inflation begins, 
it never ends and continues to create an infinite number of the so-called bubble or pocket universes.

The eternal inflation occurs in a variety of set-ups, and they can be broadly classified into three categories;
old inflation, hilltop inflation, and chaotic inflation.
The old inflation is a classic example of eternal inflation. It takes place in a false vacuum, which collapses into
one of the lower-energy vacua through tunneling and bubble formation, which is followed by slow-roll inflation.
Inside such a bubble, the inflaton energy is transferred to a hot dense plasma after the slow-roll inflation ends.
A collection of those bubble regions form pocket universes.
The entire Universe, on the other hand, continues to inflate, populating an infinite number 
of the pocket universes. Our observable Universe is contained in one of the pocket universes.

If the fundamental theory has a large number of the false vacua (e.g.~the
string landscape~\cite{Susskind:2003kw}) with each having
 different values of the coupling constants and the 
cosmological constant~\cite{Linde:1993nz,Linde:1993xx,
GarciaBellido:1993wn,Tegmark:2004qd},
 an apparently contrived fine-tuning of the parameters may be explained by anthropic arguments. 
 The eternal inflation plays a crucial role in populating various vacua in the landscape.

The eternal hilltop or chaotic inflation takes place when quantum fluctuations of the inflaton dominate
over its classical motion. In the case of eternal hilltop inflation~\cite{1983veu..conf..251S,Vilenkin:1983xq,Barenboim:2016mmw}, it occurs in the vicinity of the local maximum of the potential where the classical motion vanishes, while in the case of eternal chaotic inflation~\cite{Linde:1986fc,Linde:1986fd,Goncharov:1987ir} , it occurs at large-field 
values where the quantum fluctuation  becomes significant. In either case, once the eternal inflation happens,
it continues to inflate, which similarly helps to realize apparently fine-tuned parameters and/or initial conditions
required for e.g. the subsequent slow-roll inflation, curvaton scenarios, baryogenesis, etc.

The Swampland de Sitter Conjecture (\ref{eq:conj2})
forbids de Sitter vacua or de Sitter extrema~\cite{Obied:2018sgi}.
This immediately excludes the eternal old and hilltop inflation.\footnote{
The (quasi) de Sitter solution is subject to  
de Sitter instability from quantum backreaction. See e.g. Ref.~\cite{Matsui:2018iez}.}
In the following, therefore, we focus on the eternal inflation of chaotic type.
A typical example of the chaotic inflation is characterized by potentials of 
the form $V (\phi) \propto \phi^{\, p}$ with large field excursion 
$\Delta \phi > M_{\rm P}$~\cite{Linde:1983gd}\footnote{
Note that we consider the monomial inflaton potential during the eternal inflation,
not during the last $50 - 60$ e-folding of inflation responsible for generating 
the observed density perturbations. Indeed, 
simple chaotic inflation models with quadratic or quartic potentials are already 
ruled out by the CMB observations~\cite{Akrami:2018odb}, but this does not
apply to the eternal inflation that had occurred well before the usual slow-roll
inflation took place. 
See Refs.~\cite{Nakayama:2013jka,Nakayama:2013txa,Nakayama:2014wpa}
for polynomial chaotic inflation and its supergravity realization, which give a better fit to the CMB data~\cite{Destri:2007pv}. Also see Refs.~\cite{Linde:1983fq,Linde:1984cd,Kallosh:2007wm}.
}.
The quantum fluctuations of the inflaton on the de Sitter spacetime are frozen after
the horizon exit, and they can be considered as Brownian fluctuations or random
walks whose step size is equal to $H/2\pi$ with a time interval $H^{-1}$. 
The two-point correlation function 
$\left< { \delta\phi  }^{ 2 } \right>$ is given by
\begin{align}
\left< { \delta\phi  }^{ 2 } \right> =\frac{ { H }^{ 3 }t }{ 4{ \pi  }^{ 2 } },
\end{align}
which grows with the cosmic time. Here we approximate that the Hubble parameter is constant
for simplicity, but we take into account the dependence of the Hubble parameter on the inflaton field
in the next section.

In order to see how the eternal chaotic inflation takes place,  let us consider
a Hubble-size patch with a physical size $R\simeq H^{-1}$ where the inflation takes place.
After the time interval $t$, this region exponentially expands to the size of
$R\simeq H^{-1}\exp\left( Ht \right)$, which contains $\mathcal{N}_{\rm patch} \simeq \exp\left( 3Ht \right)$
Hubble patches. The inflaton field takes a different value in each Hubble patch due to the
accumulated quantum fluctuations. 
Let $P\left(\phi,t\right)$ denote the probability for the inflaton field to be equal to $\phi$ at the time $t$
in a Hubble patch. As an extreme case, let us estimate the probability for $\phi$ to increase
and roll up the potential all the time during the time interval $t$.
Such probability $P\left(\phi,t\right)$ is approximately given by
\begin{align}
P\left(\phi,t\right)\sim \left(\frac{1}{2}\right)^{Ht}=e^{-Ht\ln {2} }.
\end{align}
Thus, the probability exponentially decreases with the time, as expected.
However, this does not necessarily mean that the number of such Hubble patches decreases with time,
because the physical volume also increases as $\exp\left( 3Ht \right)$, which can compensate 
the exponentially small probability. Indeed, multiplying the volume factor, one obtains
\begin{align}
e^{Ht\cdot\left(3 - \ln {2} \right)}\sim e^{2.3Ht}.
\end{align}
This means that the inflaton field keeps increasing in some Hubble patches, whose number
is actually increasing due to the inflationary expansion.

In the above simplified discussion,
we have neglected the classical motion of the inflaton and assumed that inflaton dynamics is
dominated by the quantum jumps. We also did not consider its back reaction on the Hubble
parameter. However, the essentially same thing can happen in a realistic case,
as long as the quantum fluctuation is larger than the classical motion in a Hubble time.
As a result,
the inflaton can roll up the potential in some Hubble patches due to the fluctuations,
and inflation continues to take place somewhere in the Universe. Remarkably the eternal
chaotic inflation does not require the potential minimum nor maximum, and in principle
it can satisfy the swampland conjectures. We will study this issue in more detail in the next section.

%%%%%%%%%%%%%%%%%%%%%%%%%%
\section{Eternal Chaotic Inflation and Swampland}
\label{sec:eternal/swampland}
%%%%%%%%%%%%%%%%%%%%%%%%%%

In this section we study if the eternal chaotic inflation can be realized while satisfying the 
swampland conjectures. 
As we have seen before, eternal inflation takes place if
quantum fluctuations dominate over the classical variations of the inflaton field
over the Hubble time~\cite{Goncharov:1987ir,Guth:2007ng}.
Let us assume that inflation is driven by a single scalar field $\phi$.
We define the slow-roll parameter as follows,
\begin{align}
\epsilon =
\frac{M_{\rm P}^2}{2} \left(\frac{V'\left(\phi\right)}{V\left(\phi\right)}\right)^2,
\ \eta = M_{\rm P}^2\frac{V''\left(\phi\right)}{V\left(\phi\right)}.
\end{align}
If the above slow-roll parameters are much smaller than unity, the inflaton 
slowly rolls on the potential, and its dynamics 
is described by the following equations,
\begin{align}
 3 H \dot \phi \simeq - V'\left(\phi\right), \quad 
{ H }^{ 2 }\simeq\frac { V\left( \phi  \right)  }{ 3{ M }_{ \rm P }^{ 2 } },
\end{align}
where the prime denotes the derivative with respect to $\phi$.

Neglecting the inflaton mass, the quantum fluctuations are approximately written as
\begin{align}
\left\langle\delta\phi\right\rangle_{\rm q}
 \approx \frac{H}{2 \pi}.
\end{align}
On the other hand, 
the classical field variation over the Hubble time $H^{-1}$ is
\begin{align}
\left\langle\delta\phi\right\rangle_{\rm c}  
= H^{-1} |\dot{\phi}|.
\end{align}
The eternal inflation takes place if $\left\langle\delta\phi\right\rangle_{\rm q}$ is larger than
$\left\langle\delta\phi\right\rangle_{\rm c}$~\cite{Goncharov:1987ir,Guth:2007ng},
\begin{align}
\frac{\left\langle\delta\phi\right\rangle_{\rm q}}
{\left\langle\delta\phi\right\rangle_{\rm c}}
= \frac{H^2}{2 \pi |\dot\phi|} \gtrsim  1.\label{eq:eternal}
\end{align}
Note that this implies that density perturbations of the modes that exited the horizon during eternal inflation
have size of order unity, and the eternal inflation cannot be responsible for the density perturbations at 
the currently observable scales. We need another slow-roll inflation or other mechanism for generating
the observed density perturbations. 

Once the condition (\ref{eq:eternal}) is satisfied, eternal chaotic inflation takes place, 
which would populate many vacua in the landscape, if exist.
Using the slow-roll parameter $\epsilon$, one can rewrite the condition
 for eternal inflation  as
\begin{align}
\frac{ H }{M_{\rm P}}\gtrsim 2\sqrt { 2 } \pi { \epsilon  }^{ 1/2 }.
\end{align}
By using the constraint  (\ref{eq:epsilonc}) on $\epsilon$ from 
the Swampland de Sitter conjecture, we arrive at
\begin{align}
\frac{ H }{M_{\rm P}}\gtrsim 2\pi c.
\label{HMp}
\end{align}
In order for the inflaton potential not to exceed the Planck energy, $V < M_P^4$,
we also need $2\sqrt{3} \pi c < 1$.
Thus, the eternal chaotic inflation is possible only if 
the Hubble parameter is parametrically close to the Planck scale 
unless $c$ is much smaller than unity, and
quantum gravity effects may be significant. See e.g. Refs.~\cite{Buchbinder:1992rb,Starobinsky:1980te,Hawking:2000bb,Shapiro:2001rh,Salles:2014rua,Matsui:2018iez}.

For example, if we assume a mild hierarchy, $H /{M_{\rm P}} \sim 0.1$, 
to suppress possible quantum gravity effects,  the constant $c$ must be $c \lesssim  0.01$.
which is even tighter than the CMB bound, $c<0.094$ (see discussion below Eq.~(\ref{eq:epsilonc})).
This implies that, for the eternal inflation to be described by the classical treatment of the spacetime,
one of the slow-roll parameters $\epsilon$ will be much smaller than the current upper bound. 

Assuming the inflaton potential shape, one can estimate the inflaton field value where the eternal 
inflation is possible. In the case of the quartic potential, $V = \lambda \phi^4$, 
 this is given by $\phi_c \simeq 3.5 \lambda^{-1/6} M_{\rm P}$.
 By combining (\ref{HMp}), we obtain
\begin{align}
{\cal D} \gtrsim \frac{4}{\sqrt{3}c}. 
\end{align}
%leading to ${\cal D} \gtrsim 3.5 \lambda^{-1/6}$. Combined with (\ref{HMp}), we also obtain
%$\lambda \gtrsim 12.5 c^6$. Substituting the
Thus, we need $c \lesssim {\cal O}(0.01)$ and $1/{\cal D} \lesssim {\cal O}(0.01)$ 
for the eternal chaotic inflation to take place.

%%%%%%%%%%%%%%%%%%%%%%%%%%
\section{Discussion}
\label{sec:criteria2}
%%%%%%%%%%%%%%%%%%%%%%%%%%
The quantum fluctuations on the expanding universe
can bring the inflaton field back to the initial state~\cite{Linde:1986fd} or
false vacuum~\cite{Garriga:1997ef}.
Such recycling process of the inflation makes the inflation eternal.
Let us consider the possibility that the inflaton field at the end of the slow-roll inflation
rolls up the potential by quantum fluctuations, following the argument of Ref.~\cite{Linde:1993xx}.
The probability density function $P\left( \phi\right)$ of
finding the inflaton field of the value $\phi$ 
is given by a Gaussian distribution with the variance 
$\left< { \delta\phi  }^{ 2 } \right> ={ { H }^{2 } }/{ 4{ \pi  }^{ 2 } }$
for one Hubble time~\cite{Linde:1991sk}.

Now, we consider a situation where at the end of the inflation 
the inflation field $\phi_{\rm end}$ jumps up and 
becomes $\phi$ through quantum jump in some Hubble patches. 
At the end of large-field inflation, the classical field variation during the Hubble time
is of order the Planck mass, and so, the inflaton needs to jump upward by more than
the Planck mass to overcome the classical motion. 
Let us denote the quantum jump by $C\cdot M_{\rm P}$ with $C \gtrsim 1$. Then 
the probability for such a large quantum jump is given by
\begin{align*}
P\left( \phi_{\rm end} \rightarrow  \phi_{\rm end} + C M_{\rm P} \right)
&\approx
\exp \left( \frac {-2\pi^2C^2 M_{\rm P}^{ 2 } }{ H^{ 2 }_{\rm end} }  \right),
\end{align*}
where we have used the gaussian distribution with the variance $\left< { \delta\phi  }^{ 2 } \right> ={ { H }^{2 } }/{ 4{ \pi  }^{ 2 } }$.
Suppose that, due to the previous inflationary expansion, 
 there is a large number of the Hubble patches just before the inflaton
experienced the large jump. 
The number of the Hubble patches that experience such a large jump is estimated as
\begin{align}
\mathcal{N}_{\rm patch}\approx e^{3N_e }
P\left( \phi\right) \approx \exp \left(
3N_e - \frac {2\pi^2C^2 M_{\rm P}^{ 2 } }{ H^{ 2 }_{\rm end} } \right) ,
\end{align}
where $N_e$ is the total e-folding number of the previous inflationary expansion. 
Thus, if we use $C \approx 1$, we obtain
\begin{align}
N_e \gtrsim \frac{\pi^2}{3}\frac{M_{\rm P}^2}{H_{\rm end}^2} ,
\end{align}
to find more than one such Hubble patch. In order to goes back
to the inflaton field value where the eternal inflation takes place,
the total e-folding number must be even larger. 

The required lower bound on the e-folding number is in tension
with the Swampland distance conjecture since it heavily restricts 
the duration of the inflation~\cite{Agrawal:2018own,
Achucarro:2018vey,Dias:2018ngv}. 
By using the Swampland distance
criterion and the Lyth bound~\cite{Lyth:1996im} 
we connect the field variance $\Delta \phi$ and 
the total e-folding number $N_e$ as follows:
\begin{align}
N_e \sqrt{2\epsilon}  \lesssim \dfrac{\Delta \phi}{M_{\rm P}} 
\quad\Rightarrow\quad 
60 \lesssim N_e  \lesssim \mathcal{D}/c\label{eq:EF}\, ,
\end{align}
where the lower bound of Eq.~(\ref{eq:EF}) comes from the 
cosmological observations to solve the horizon and flatness problems.
Thus, the recycling process of the eternal inflation is also restricted by 
this bound.

%%%%%%%%%%%%%%%%%%%%%%%%%%
\section{Conclusions}
\label{sec:discussion}
%%%%%%%%%%%%%%%%%%%%%%%%%%
The eternal inflation is believed to play an important role in populating various vacua in the 
landscape if they exist. It also helps to realize apparently fine-tuned parameters or set-ups based
on the anthropic arguments. In this paper
we have studied if the eternal inflation can be realized while satisfying the recently proposed
Swampland conjectures. We found that only the eternal inflation of chaotic type is possible.
The Hubble parameter during the eternal inflation is parametrically close to the Planck scale,
and it requires the numerical constants $c$ and $1/{\cal D}$ appearing in the conjectures 
to be of ${\cal O}(0.01)$ or smaller. Even though the eternal inflation is not responsible for the
observed density perturbations, the derived constraints on the numerical coefficients 
$c$ and $1/{\cal D}$ have interesting implications for observables such as the tensor-to-scalar
ratio as well as the dark energy equation of state through the swampland conjectures. 

\vspace{3mm}
%%%%%%%%%%%%%%%%%%%%
\para{Acknowledgements} 
%%%%%%%%%%%%%%%%%%%%
F.T.  thanks Prateek Agrawal for useful communications. 
F.T. thanks the hospitality of MIT Center for Theoretical Physics where this work was done. 
This work is partially supported by JSPS KAKENHI Grant Numbers JP15H05889 (F.T.), 
JP15K21733 (F.T.), JP17H02878 (F.T.), and JP17H02875 (H.M. and F.T.), Leading Young Researcher Overseas
Visit Program at Tohoku University (F.T.), and by World Premier International Research Center
 Initiative (WPI Initiative), MEXT, Japan (F.T.).

%%%%%%%%%%%%%%%%%%%%
%%%%%%%%%%%%%%%%%%%%
\bibliography{references}
%%%%%%%%%%%%%%%%%%%%
%%%%%%%%%%%%%%%%%%%%

\end{document}